\documentstyle[preprint,prl,aps]{revtex}
\begin{document}
\draft

\preprint{\vbox{\hbox{March 1996 }\hbox{rev. May 1996}\hbox{IFP-725-UNC}}}

\title{Horizontal Symmetry for Quark and Squark Masses in Supersymmetric SU(5).}
\author{\bf Paul H. Frampton and Otto C. W. Kong}
\address{Institute of Field Physics, Department of Physics and Astronomy,\\
University of North Carolina, Chapel Hill, NC  27599-3255}
\maketitle

\begin{abstract}
Recent interest in horizontal symmetry model building has been driven mainly by the large top mass and hence strong hierarchy in quark masses, and the possibility of appropriately constrained soft squark mass matrices,  
in place of an assumed universality condition, for satisfying the relevant FCNC constraints. Here we 
present the first successful SUSY-$SU(5)$ model that has such a feature. The horizontal symmetry is a 
gauged $(Q_{12} \times U(1))_H$ ($\subset (SU(2) \times U(1))_H$). All nonrenormalizable terms 
compatible with the symmetry are allowed in the mass matrix constructions. Charged lepton masses 
can also be accommodated.  

\end{abstract} 
\pacs{} 

\newpage 

{\it Introduction.} 
Despite the success of the Standard Model (SM) and the very encouraging 
indication of its plausible supersymmetric unification (SUSY-GUT), we 
still lack a real understanding of flavor physics. In this perspective,
the idea of a horizontal (flavor/family) symmetry has been resurrected 
as the most popular candidate theory to supplement the vertical 
(unified) gauge theory of particle physics. Various authors have 
illustrated the interesting model-building possibilities in using 
spontaneously broken horizontal symmetry to constrain the Yukawa 
sector of the SM with the aim at obtaining phenomenologically-viable
texture patterns for the quark mass matrices\cite{u1,LNS,PS,KS,q2n,q2n2,552}.
The authors of this letter have concentrated on the more
restrictive scenario of a gauged nonabelian horizontal symmetry,
$SU(2)$ and its discrete dicyclic subgroups $Q_{2N}$\cite{q2n},
which is compatible with vertical unification\cite{q2n2,552}.

While there is quite a list of interesting extended applications of
a horizontal symmetry, the most interesting one is
no doubt its use in constraining squark mediated FCNC in a SM
supplemented with softly broken supersymetry, which is favored by
the unification picture. Any horizontal symmetry on the low
energy fermions naturally constrains (soft) couplings among their
SUSY-partners. In fact, the use of a horizontal symmetry in the place
of an imposed degeneracy among squark masses is one of the major
motivation in the recent resurrection of the theory\cite{qsa,sd}.
A $SU(2)$ (or $U(2)$)
horizontal symmetry with the lighter two families forming a 
doublet has then been advocated by some authors\cite{PS,q2n,552,sd,PT}. 
In this letter, we will present the first successful model, with
a $(Q_{12}\otimes U(1))_H$ ($\subset (SU(2)\otimes U(1)_H$)
horizontal symmetry compatible with a vertical SUSY-$SU(5)$ unification.

{\it The FCNC Constraints.}
Before going into the model-building specifics, we summarize
below the relevant background concerning the squark mediated
FCNC in neutral meson mixings\cite{fcnc}.

The $6 \times 6$ squark mass-squared matrices $\tilde{M}^{u2}$ and
$\tilde{M}^{d2}$ are each divided into four $3\times 3$ sub-matrices.
The leading contributions to the off-diagonal blocks, $\tilde{M}^{u2}_{LR}$
and $\tilde{M}^{d2}_{LR}$, arise from
the trilinear $A$-terms, while the leading contributions to the 
diagonal blocks, $\tilde{M}^{u2}_{LL}, \tilde{M}^{u2}_{RR}, 
\tilde{M}^{d2}_{LL}$ and $\tilde{M}^{d2}_{RR}$,
arise from the soft mass terms. The latter dominate over the former,
and can generally lead to unacceptably large FCNC-effect in neutral meson mixing
when universality of soft masses is not imposed. The flavor changing
quark-squark-gluino couplings are the result of the fact that a generic
squark mass-squared matrix cannot be simultaneously diagonalized 
with the corresponding quark mass matrix. For
instance, constraints from $K-\bar{K}$ and $B-\bar{B}$ mixing on 
$\tilde{M}^{d2}_{LL}$ can be expressed by an upper bound on
\begin{equation}
(\delta ^d_{LL})_{12} = \frac{1}{\tilde{m}^2} (\tilde{m}^2 _1 K_{11} K^{\dag}_{12} +  \tilde{m}^2 _2
 K_{12}  K^{\dag}_{22} + \tilde{m}^2 _3  K_{13}  K^{\dag}_{32})
\end{equation}
and
\begin{equation}
(\delta ^d_{LL})_{13} = \frac{1}{\tilde{m}^2} (\tilde{m}^2 _1 K_{11} K^{\dag}_{13} +  \tilde{m}^2 _2
 K_{12}  K^{\dag}_{23} + \tilde{m}^2 _3  K_{13}  K^{\dag}_{33})
\end{equation}
respectively, where $\tilde{m}^2 _i$ are the three eigenvalues and 
$\tilde{m}^2$ their average, and $K$ is
actually $K^d_L = V^d_L\tilde{V}^{d\dag}_L$ with $\tilde{V}^{d}_L$ being
the unitary matrix that diagonalize $\tilde{M}^{d2}_{LL}$
and $V^d_L$ the usual notation for the matrix involved in diagonalizing quark
masses.
There are also constraints on the respective elements of
$\delta ^d_{RR}$, and mixed product of the form
$\left \langle \delta ^{d}_{ij}\right \rangle= ( (\delta ^d_{LL})_{ij}
(\delta ^d_{RR})_{ij} )^{1/2}$. There are similar constraints from
$D-\bar{D}$ mixing on the corresponding up-sector quantities. 
While the actual  numerical 
bounds depend on the details of the SUSY-spectrum, an illustrative
set of numbers are listed in Table 1.

In principle there are other very important
flavor-changing processes, such as $b \longrightarrow s \gamma$\cite{bsg},
that constrain the off-diagonal blocks, $\tilde{M}^{d2}_{LR}$.   
However, while universality of squark masses is not a natural
consequence of horizontal symmetry, proportionality of the trilinear
soft $A$-terms to the quark Yukawa couplings could be, provided that
the horizontal symmetry is not an $R$-symmetry. This then
would take care of the necessary FCNC suppression arising from
$\tilde{M}^{u2}_{LR}$ and $\tilde{M}^{d2}_{LR}$. Hence, we are not going
to discuss the off-diagonal blocks any further.

Another important question involved is the scale where any structure on
the squark masses is imposed.
On the one hand, it is possible to  have  universality among
the soft SUSY-breaking terms imposed at the Planck-scale 
yet significantly corrected at the GUT-scale\cite{pgs,pgs1,pgs2}
leading to interesting lepton flavor violating and CP-violating signal\cite{pgs2}.
On the other hand, there is the scenario where non-universal
squark masses are rendered sufficently degenerate by large common
contributions from RG-evolution due to particularly heavy gauginos\cite{gcs}.
Scenarios of this second type are also possible in some string-inspired supergravity
models\cite{sgc}.

For our model-building consideration, we are interested only in constraints 
on non-universal squark masses which result from a horizontal symmetry
spontaneously broken at some high energy scale. A recent analysis by
Choudhury et.al.\cite{Cea} in the MSSM framework is most relevant. The result can be summarized
by three points: {\it 1)} large gauginos masses enhance the diagonal
squark masses; {\it 2)} non-universal $A$-terms decrease the off-diagonal
mass-squared matrix elements;{\it 3)} this $A$-term suppression effect
decreases as the top Yukawa gets large and approaches zero 
at its IR (quasi-)fixed point. We then conclude that for a horizontal
symmetry model with a hierarchical quark mass texture, it is sufficient
for  the FCNC constraints to be 
satisfied naively by the high energy texture of the squark mass-squared
matrices ($\tilde{M}^{u2}_{LL}, \tilde{M}^{u2}_{RR}, 
\tilde{M}^{d2}_{LL}$ and $\tilde{M}^{d2}_{RR}$); and
in the absence of very massive gauginos. 
the necessary FCNC bounds are not going to be very much weakened at
the high scale\cite{rge}. We aim at providing such a model with the FCNC constraints
satisfied by the squark mass-squared texture from a broken horizontal symmetry,
the energy scale of which to be specified later.

{\it $2+1$ Family Structure in SUSY-GUT.} 
Consider in the SUSY-$SU(5)$ framework a general $2+1$ family structure.
We label the chiral supermultiplets that contain the low energy chiral
fermions as, for the doublets containing the first and second families, 
$10_2$ and $\bar{5}_2$, and for the third family singlets,
$10_1$ and $\bar{5}_1$; and  $5$ and $\bar{5}$ represent the Higgses. 
 We want only the top  quark 
to have a mass term invariant under the horizontal symmetry. We can
take both the $10_1$ and the $5$ to be in representation ($1,0$)
of the $(Q_{2N}\otimes U(1))_H$,
where the zero $U(1)$ charge is taken for simplicity. Denote the
representation of the $10_2$ by ($2_k, A$), where $2_k$ is a general doublet
of a $Q_{2N}$ and $A$ the $U(1)$ charge. The nontrivial $U(1)$ charge is what
forbids an invariant mass for the doublet. Now, if we take a $SU(5)$ singlet $\phi$
in ($2_k, -A$), with a horizontal symmetry breaking VEV in the direction
$[1,1]$\cite{t3} of the doublet. We have from the terms
\[
10_2 10_1 \left \langle 5\right \rangle\left \langle \phi \right \rangle_{sym} / M_{Pl} ,
\hspace{.5cm} 
10_2 10_2 \left \langle 5\right \rangle\left \langle \phi \right \rangle_{sym}^2 / M_{Pl}^2 ,
\]
($M_{Pl} \sim 2.4\times 10^{18} GeV$) an up-quark mass matrix of the form
\begin{equation}
$$M_{u} \sim \left(\begin{array}{ccc} 
\lambda^{4} & \lambda^{4} & \lambda^{2} \\
\lambda^{4} & \lambda^{4} & \lambda^{2} \\
\lambda^{2} & \lambda^{2} & 1
\end{array}\right)$$ ,
\end{equation}
where $\lambda \sim .22$, coefficients of order one are neglected, and we set 
\begin{equation}
\left \langle \phi \right \rangle_{sym} / M_{Pl} \sim  \lambda^{2} ,
\end{equation}
The VEV $\left \langle \phi \right \rangle_{sym}$ together with its conjugate also give us 
off-diagonal terms in $\tilde{M}^{2}_{10}$, through similar higher 
dimensional terms, as
\begin{equation}
$$\tilde{M}^{2}_{10} \sim \left(\begin{array}{ccc} 
1 & \lambda^{4} & \lambda^{2} \\
\lambda^{4} & 1 & \lambda^{2} \\
\lambda^{2} & \lambda^{2} & 1
\end{array}\right)$$ .
\end{equation}
If we put in another VEV for $\phi$ (denoted by $\left \langle \phi \right \rangle_{antisym}$) in the
$[1,-1]$ direction of the doublet, with  
\begin{equation}
\left \langle \phi \right \rangle_{antisym} / M_{Pl} \sim  \lambda^{4} ,
\end{equation}
this gives nonzero mass to the up. We have then a mass matrix of the form
\begin{equation}
$$M_u = \left(\begin{array}{ccc} 
a + x & a & (c+y) \\
a & a - x & (c-y) \\
(c+y) & (c-y) & 1
\end{array}\right) $$
\end{equation}
where
\begin{equation} 
a \sim \lambda^{4}, \ \ \ x \sim \lambda^{6}, 
\ \ \ c \sim \lambda^{2}, \ \ \ y \sim \lambda^{4}.
\end{equation}
There are also extra contributions to $\tilde{M}^{2}_{10}$ of higher order 
in $\lambda$ that we neglect.

The choice of scales for the VEVs of $\phi$ are consistent. The two VEVs
correspond to two linear independent states of the $2_k$ doublet. If
$Q_{2N}$ breaks to a $Z_2$ remnant at $\lambda^{2}M_{Pl}$, with the $[1,1]$
state from the doublet transform trivially under $Z_2$ and the $[1,-1]$
state transform  non-trivially, the latter VEV  would be further
suppressed till the breaking of the $Z_2$ remnant. So, in the hierarchical
basis, the $Z_2$ symmetry protects the first family, the $u$ quark,
from getting a mass; in the horizontal symmetry basis considered here, 
it enforces the degeneracy between the lighter two families.

Note that $\tilde{M}^{2}_{10}$ contains $\tilde{M}^{u2}_{LL},
\tilde{M}^{u2}_{RR}$ and $\tilde{M}^{d2}_{LL}$ which  share the
same texture pattern of the parent $\tilde{M}^{2}_{10}$. 
We have, then, through introducing the two $\phi$ VEVs, an $M_u$,
which corresponds to a acceptable symmetric texture pattern,
and an $\tilde{M}^{2}_{10}$ that satisfy all the correspondent constraints,
when a compatible down-quark mass matrix is assumed. The great economy
of the scheme is self-evident. 

We leave the details concerning the admissible texture patterns for 
quark and squark masses in  the half-democratic 
half-hierarchical form given above to a separate publication\cite{qst}.

{\it Gauge Anomaly Cancellation.}
Before presenting our complete model,
we comment on the gauge anomaly cancellations.
We have a gauged $SU(5)\otimes (Q_{2N}\otimes U(1))_H$ symmetry ($N=6$ in particular),
with $U(1)$ being replacable by a $Z_N$ subgroup.
The first thing to notice is that all chiral
supermultiplets have to be embeddable into complete $SU(5)\otimes SU(2)$ 
representations, to be free from any anomalies involving only $SU(5)$ and $SU(2)$. 
This is a nontrivial condition, making the situation different from gauging
abelian discrete symmetries\cite{gzn}. In our model, for example, 
we take a $10$ and a $\bar{10}$ from a $4$ and $2$ of
 $SU(2)$ respectively, assuming conjugate $U(1)$ charges. Breaking the $SU(2)$ 
 to the discrete $Q_{12}$ (or any $Q_{2N}$ with $N\geq 4$)
subgroup, we have the splitting
\[ 4 \longrightarrow 2_3 + 2_1, \ \ \ \ \ 2 \longrightarrow 2_1. \]
A $Q_{12}$ invariant Dirac mass term can develop for the $2_1$ doublet,
leaving behind a chiral $(10,2_3)$, to be identified as our $10_2$.

We assume  that the supermultiplets containing the quarks and leptons are
the only chiral content, with all the other multiplets in matching vector-like
pairs. The latter are naturally heavy, except the EW-breaking Higgs doublets.
Cancellation of the $[SU(5)]^2U(1)$  anomaly has to be enforced.
The  situation for the $[SU(2)]^2U(1)$ and $U(1)$ anomalies 
is, however, more like the abelian scenario. It is
possible, for example, to introduce extra $SU(5)$ singlet supermultiplets that
can develop Dirac or Majorana masses invariant under $Q_{12}\otimes Z_N$. 

{\it The Full $(Q_{12}\otimes U(1))_H$ Model.}
Along the lines considered above,  it is possible to build a full model
which has a gauged horizontal symmetry that accounts for both the quark and squark 
mass matrix textures and fits all the phenomenological constraints. Here
we present the example which we believe to be the most economic. 
It remains to be seen whether the assumed sequence of horizontal
symmetry breaking can be naturally obtained from a scalar potential.

We have $10_1$ and the Higgs multiplets $5$ and $\bar{5}$ in $(1,0)$,
and  the $10_2$ horizontal doublet in a $(2_3, 1)$ of $(Q_{12}\otimes U(1))_H$
as mentioned above. We further put the $\bar{5}_1 + \bar{5}_2$  
in a $SU(2)$ triplet, which then splits into a $1^{'} + 2_2$ at the
$Q_{12}$ level. The full representation assignments
of the chiral supermltiplets are shown in Table 2. Now we can simply take
the above results on $M_u$ and $\tilde{M}^{2}_{10}$, setting $k=3$ and
$A=1$. To complete the model for $M_d$ and $\tilde{M}^{2}_{\bar{5}}$, we need
a few extra heavy VEVs, as given in Table 2. Tracking down all the lower 
order coupling (up to $\lambda ^{6}$), we obtain 
\begin{equation}
$$M_{d} \sim \lambda^{2} \left(\begin{array}{ccc} 
a' + x' & a' & c' + y' \\
a' & a' - x' & c' - y' \\
z' & z' & 1
\end{array}\right)$$ ,
\end{equation}
where 
\begin{equation} 
a' \sim \lambda^{3}, \ \ \ x' \sim \lambda^{4}, 
\ \ \ c' \sim \lambda^{2}, \ \ \ y' \sim \lambda^{4}, \ \ \ z' \sim \lambda^{3};
\end{equation}
and for the squarks mass matrix $\tilde{M}^{d2}_{RR}$ from
\begin{equation}
$$\tilde{M}^{2}_{\bar{5}} \sim \left(\begin{array}{ccc} 
1 & \lambda^{4} & \lambda^{3} \\
\lambda^{4} & 1 & \lambda^{3} \\
\lambda^{3} & \lambda^{3} & 1
\end{array}\right)$$ .
\end{equation}
For example,
\begin{equation}
a' \lambda ^{2} \sim \left \langle\bar{5} \right \rangle\left \langle(2_2,-1)\right \rangle_{sym} \left \langle(2_3,1)\right \rangle_{sym} \left \langle(1,1)\right \rangle_{sym} / M_{Pl}^{3} ,
\end{equation}
\begin{equation}
x' \lambda ^{2} \sim \left \langle\bar{5} \right \rangle\left \langle(2_5,-1)\right \rangle_{antisym} \left \langle(1',2)\right \rangle/ M_{Pl}^{2} .
\end{equation}

Note that $M_d$ is (slightly) not symmetric. This is a general feature of $SU(5)$
unification. The asymmetric mass matrix can be put into a symmetric
form by a rotation of the right-handed down-quark field, raising
the $31-$and $32-$ entries to the same order as the $13-$ and $23-$ ones, as noted by
the authors previously\cite{552}. The extra rotation makes $V_R^d$
different from $V_L^d$, and is relevant to the constraints on $\tilde{M}^{d2}_{RR}$.
Detail analysis shows that this actually leads to slight further suppression of the 
$13-$ and $23-$ entries in $K^d_R$. The results concerning the FCNC constraints are
shown in Table 1.

To accommodate the charged lepton masses, either the Georgi-Jarlskog\cite{GJ}
or the Ellis-Gaillard\cite{EG} mechanism can be used. While there may be potentially
complications and interesting phenomenology involved\cite{pgs2}, we will leave
the detail features of the leptonic sector for future investigation. 

We note also that
there is the possibility of obtaining the gravitationally induced nonrenormalizable
terms through a Froggatt-Nielsen\cite{FN} mechanism thereby reducing the
horizontal symmetry breaking scale.  

Finally, we want to point out that the model has not addressed the 
doublet-triplet splitting problem. One can assume the simple fine-tuning
solution. Apart from its being "unnatural", there is also an extra recent
objection from the perspective of precise gauge coupling unification.
The later problem can however be corrected by some other strategy\cite{ARU}.
"Missing doublet" models provides a very interesting alternative that is free
from both problems\cite{gcf}, as well as giving less unnatural mass constraints
for an acceptable proton decay rate\cite{pd}. 
Extensions or modifications of the model to
incorporate  a missing doublet structure and a suppression of squark-mediated
proton decay without R-parity  are under investigation.

This work was supported in part by the U.S. Department of 
Energy under Grant DE-FG05-85ER-40219, Task B.\\


\newpage

\bigskip

\begin{center}

\begin{tabular}{c|cccc}\hline\hline
$K-\bar{K}$ mixing	& $(\delta ^d_{LL})_{12}$ 	& $(\delta ^d_{RR})_{12}$	& $\left \langle\delta ^{d}_{12}\right \rangle$\\ \hline
upper bound		& $0.05$ & $0.05$ & $0.006$\\ \hline
our model		& $\sim \lambda ^{5}$ & $\sim \lambda ^{5}$ & $\sim \lambda ^{5}$	\\ \hline \hline
$B-\bar{B}$ mixing	& $(\delta ^d_{LL})_{13}$	& $(\delta ^d_{RR})_{13}$	& $\left \langle\delta ^{d}_{13}\right \rangle$\\ \hline
upper bound		& $0.1$  & $0.1$  & $0.04$  \\ \hline
our model		& $\sim \lambda ^{3}$ & $\sim  \lambda ^{4}$ & $\sim \lambda ^{3.5}$	\\ \hline \hline
$D-\bar{D}$ mixing	& $(\delta ^u_{LL})_{12}$	& $(\delta ^u_{RR})_{12}$	& $\left \langle\delta ^{u}_{12}\right \rangle$\\ \hline
upper bound		& $0.1$  & $0.1$  & $0.04$   \\ \hline
our model		& $\sim \lambda ^{6}$ & $\sim \lambda ^{6}$ & $\sim \lambda ^{6}$ 	\\ \hline \hline
\end{tabular}
\end{center}

\vspace{0.1in}
Table 1: Constraints from neutral meson mixings and results of our
model.

\vspace{2in}

\noindent
\begin{tabular}{l|cccc|l}\hline\hline
$SU(5)$ multiplet 		& $10_2 $	& $10_1$	& $\bar{5}_2$	& $\bar{5}_1$	& $5 + \bar{5}$ \\ \hline
$(Q_{12}\otimes U(1))_H$ rep. 	& $(2_3,1)$	& $(1,0)$	& $(2_2,-2)$	& $(1^{'},-2)$	& $(1,0)$	\\
\hline\hline
\end{tabular}
\begin{tabular}{r|lrl}
\multicolumn{4}{l}{$SU(5)$ singlet heavy VEVs -- their	$(Q_{12}\otimes U(1))_H$ rep.} \\ \hline
---\ \ $\left \langle\phi _i \right \rangle_{sym}$  $\sim \lambda ^{2}M_{Pl}$	
& $(2_3,-1)$	& $(2_2,-1)$	& $(2_3,1)$	 \\
---\ \ $\left \langle\phi _i \right \rangle_{antisym}$  $\sim \lambda ^{4}M_{Pl}$	
& $(2_3,-1)$	& $(2_5,-1)$	&		\\
---\ \  $\left \langle\phi _i \right \rangle$ 
& \multicolumn{3}{l} {$(1^{'},2) \sim \lambda ^{2}M_{Pl}$ \ $(1,1) \sim \lambda M_{Pl}$} \\ \hline\hline
\end{tabular}

\vspace{0.1in}
Table 2: Supermultiplet and heavy VEV content of our model.
The $SU(5)$ VEVs should correspond to scalar states of complete supermultiplets
in vector-like pairs and with heavy masses, for instance Planck scale masses. 
Note that for the horizontal doublets VEVs,  $\left \langle\phi _i \right \rangle_{sym}$ are in the 
$[1,1]$ direction while  $\left \langle\phi _i \right \rangle_{antisym}$ are in the $[1,-1]$ direction.
 $\left \langle\phi _i \right \rangle$ corresponds to a VEV for a $Q_{12}$ 
singlet. Notice that there are two different singlets for any $Q_{2N}$ group,
a $1^{'}$ and a $1$. Only the later is truely $Q_{2N}$ invariant. Hence the
$(1,1)$ VEV breaks only the $U(1)_H$ but not the $(Q_{12})_H$ symmetry.

\newpage



\bigskip

{\bf Table Caption.}\\

Table 1: Constraints from neutral meson mixings and results of our model.
The numerical bound are given as an illustrative set of values (from
Ref.~\cite{qsa}), details of  which depends on gaugino and squark masses.

\bigskip

Table 2: Supermultiplet and heavy VEV content of our model.
The $SU(5)$ VEVs should correspond to scalar states of complete supermultiplets
in vector-like pairs and with heavy masses, for instance Planck scale masses. 
Note that for the horizontal doublets VEVs,  $\left \langle\phi _i \right \rangle_{sym}$ are in the 
$[1,1]$ direction while  $\left \langle\phi _i \right \rangle_{antisym}$ are in the $[1,-1]$ direction.
 $\left \langle\phi _i \right \rangle$ corresponds to a VEV for a $Q_{12}$ 
singlet. Notice that there are two different singlets for any $Q_{2N}$ group,
a $1^{'}$ and a $1$. Only the later is truely $Q_{2N}$ invariant. Hence the
$(1,1)$ VEV breaks only the $U(1)_H$ but not the $(Q_{12})_H$ symmetry.


\begin{references}
\bibitem{rge}
Much in the same token, the RG-running effects are generally not directly addressed
in horizontal symmetry model building, see for instance Refs.~\cite{LNS,PS,PT}.
\end{references}
\end{document}